\documentclass[review]{elsarticle}

\usepackage{lineno,hyperref}
\usepackage{mathpazo}

\modulolinenumbers[5]
\usepackage[a4paper]{geometry}
\usepackage{textcomp}
\usepackage{graphicx}
\usepackage{color}
\usepackage[scanall]{psfrag}
\def\figfont{\footnotesize}

\newcommand{\RTO}{$\mbox{Rb}_{2}\mbox{Ti}_2\mbox{O}_5$}

\def\be{\begin{equation}}
\def\ee{\end{equation}}
\def\bea{\begin{eqnarray}}
\def\eea{\end{eqnarray}}

\begin{document}

\begin{frontmatter}

\title{Virtual cathode induced in $\mbox{Rb}_{2}\mbox{Ti}_{2}\mbox{O}_{5}$ solid electrolyte}
\author{Sofia de Sousa Coutinho, Rémi Federicci, Brigitte Leridon* and Stéphane Holé} 
\address{LPEM --  ESPCI Paris -- PSL University -- Sorbonne Université -- CNRS\\ 10 rue Vauquelin -- 75005 Paris -- France}



\cortext[Brigitte Leridon]{Corresponding Author}
\ead{brigitte.leridon@espci.fr}


\begin{abstract}
\RTO\ (RTO) has recently been demonstrated to be a solid electrolyte, producing colossal capacitance when interfaced with metals. In order to understand the mechanisms leading to such colossal equivalent permittivity (up to four orders of magnitude above state-of-the-art values), the charge distribution in RTO is a key feature to be investigated. In the present article, this charge distribution is probed using the pressure-wave-propagation method, in devices made of RTO single crystals or polycrystals sandwiched between two metallic electrodes.  Remarkably enough, in both types of samples, negative charges are found to accumulate inside RTO, near the anode, while the electric field near the cathode remains zero. This proves that the ionic carriers are majoritarily negatively charged and provides an explanation for the colossal capacitance. The latter takes place only at the anode while the cathode is virtually shifted into the solid electrolyte.
\end{abstract}

\begin{keyword}
\texttt Ion conductor, Solid electrolyte, Rubidium titanate, Virtual cathode, supercapacitor material
\MSC[2010] 00-01\sep  99-00
\end{keyword}

\end{frontmatter}


\section {Introduction}

Solid electrolytes have attracted much attention during the last decade
in particular in the search for solid oxide fuel cells \cite{SmithaJMS2005},
and high capacitance devices \cite{VangariJEE2013,LiAEM2015}. When
a voltage is applied to these high capacitance structures, a thin
double layer of charges is formed under voltage at the interface between
the electrodes, which are electron conductors, and the solid electrolyte,
which is a purely ionic conductor. Taking advantage of the ultra-thin character
of the double layer, and of the porosity of the materials which allows
very large electrode equivalent area, it is possible to produce super-capacitors of EDLC (Electric double-layer capacitors) type. When an additional "blocking" mechanism is present such as an electrochemical reaction or an adsorption or intercalation effect, these devices are termed pseudocapacitors and take advantage of cumulated EDLC and pseudocapacitive effects \cite{augustyn_pseudocapacitive_2014}.
The use of such components in high power electrical applications,
such as for instance electric vehicles, is growing rapidly \cite{CarterTVT2012}.

The perovskite-derived material $\mbox{Rb}_{2}\mbox{Ti}_{2}\mbox{O}_{5}$ (RTO), belonging to the $\mbox{M}_{2}\mbox{Ti}_{2}\mbox{O}_{5}$ family, which structure was identified in the 1960' \cite{ANDERSSON:1960fi, Andersson:1961vf}, has been recently found to display very interesting dielectric properties,
showing an equivalent relative permittivity up to $10^{9}$ at room
temperature \cite{FedericciPRM2017}. This is three to four orders of magnitude
above competitive materials and was attributed to ionic motion.  In addition, this material has been demonstrated to exhibit memory effects also related to ionic motion \cite{FedericciJAP2018}.For obtaining such high equivalent permittivity,
it is necessary that the charges accumulate either (1) at both electrode
interfaces or (2) at one interface only, the other interface exhibiting
an Ohmic contact. In the first case, the structure is equivalent to
two capacitors in series and at least two mobile ionic species of
opposite polarity should coexist in the electrolyte, the overall equivalent
capacitance being controlled by the smaller of the two capacitances.
In the second case, the structure is equivalent to a single capacitor
and may involve only one type of mobile ionic species in the electrolyte. Previous work has been performed to calculate the charge distribution in the system under various hypotheses \cite{FedericciJAP2018}. However, if these calculations enabled to simulate the observed I-V curves, they failed at reproducing them in a quantitative way, with reasonable physical parameters, which was interpreted as pointing to a missing ingredient in the problem \cite{FedericciJAP2018}. 

The present paper aims at identifying this missing ingredient, by investigating the charge distribution inside the samples. As a matter of fact, an extremely high density of charges is required at
the interface to account for such a colossal equivalent capacitance (over 10$^5$ F g$^{-1}$).
Therefore we have used a space charge distribution measurement method,
namely the pressure-wave-propagation method developed in our Laboratory
\cite{LaurenceauPRL77}, to follow the charge buildup 
under a continuous voltage bias \cite{HoleTDEI2003}. 

\section{Materials and methods}

\subsection{Sample preparation}

\begin{figure}

\centering{}
\includegraphics[width=0.3\linewidth]{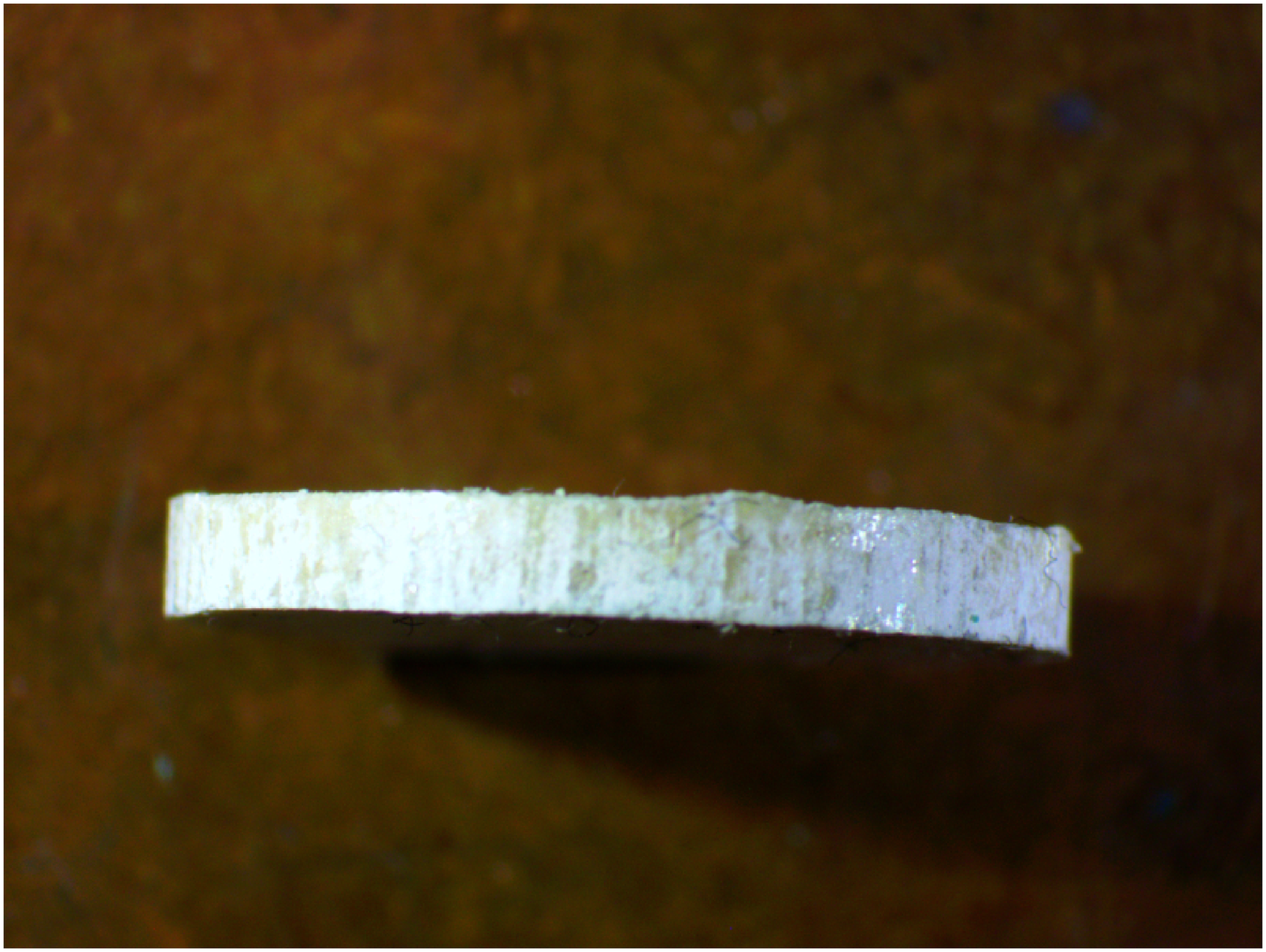}
\includegraphics[width=0.3\linewidth]{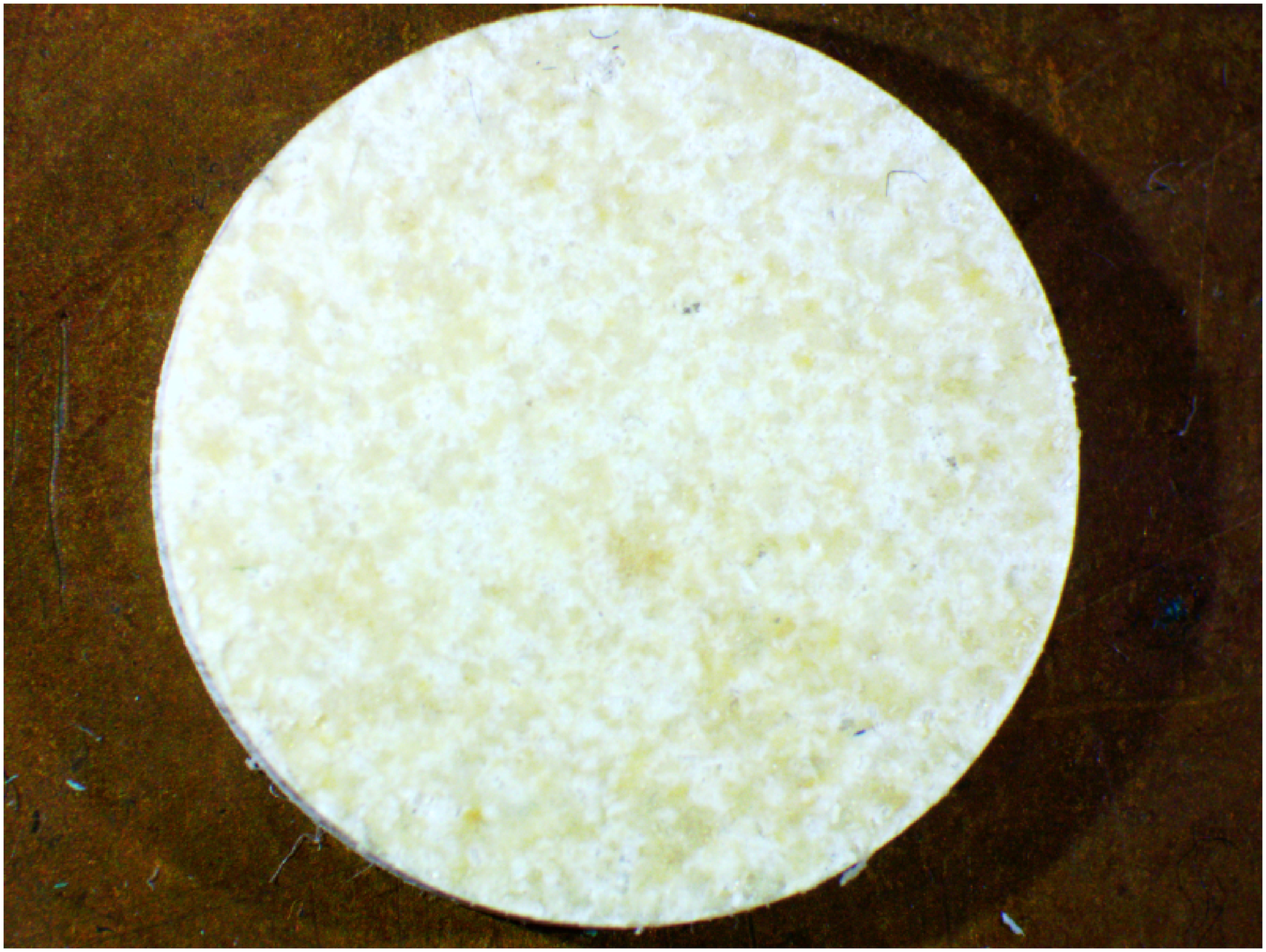}
\caption{ Photographs of a polycrystalline RTO pellet before electrode deposition. The diameter is 15mm. The differences in color are presumably due to different hydration degrees of RTO.} 
\end{figure}

\begin{figure}
\centering{}
\figfont
\psfrag{a}[Br][Br]{(a)}
\psfrag{5mm}[Bc][Bc]{5\,mm}
\psfrag{1mm}[Bc][Bc]{1\,mm}
\psfrag{Al}[Bc][Bc]{Al}
\psfrag{SRb2Ti2O5}[Br][Br]{$\mbox{Rb}_2\mbox{Ti}_2\mbox{O}_5$ crystal}
\psfrag{Epoxy}[Br][Br]{Epoxy resin}
\psfrag{caxis}[Bl][Bl]{$c$}
\psfrag{Carbon glue}[Br][Br]{Carbon glue}
\psfrag{b}[Br][Br]{(b)}
\psfrag{1.86mm}[Bc][Bc]{1.86\,mm}
\psfrag{Deposited Au}[Br][Br]{Deposited Au}
\psfrag{PRb2Ti2O5}[Br][Br]{$\mbox{Rb}_2\mbox{Ti}_2\mbox{O}_5$ ceramic}
\psfrag{Au foil}[Br][Br]{Au foil}
\includegraphics[width=0.8\linewidth]{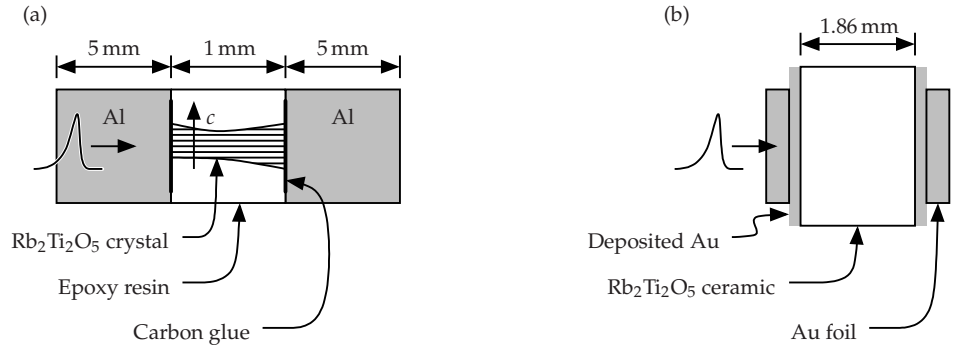}
\caption{(a) Schematics of the single crystal sample structure. (b) Schematics of the polycrystalline sample structure. } \label{fig:Sample}
\end{figure}

RTO crystals were synthesized
\cite{FedericciPRM2017,FedericciACSB2017}, and then annealed at 150 $\deg$ C
for 24h under 5~mbar of nitrogen gas. Then, to fabricate samples, one single crystal per sample was selected and embedded in Epoxy resin (Araldite D). After polymerization at 50$\deg$ C, resin and crystal were cut in 1-mm thick slices in such a way that the crystalline $ab$-planes lay perpendicular to the surface of the
sample. Finally two 5-mm-thick aluminum electrodes were glued on both
sides with carbon paste. This ensured a good mechanical coupling as
well as a stable sample structure for carrying out all the measurements.
The full sample was 15-mm in diameter but the RTO
single crystal cross section was typically 1~mm$\times$1.5~mm.
The structure of a typical single-crystal-type sample is described in Figure~\ref{fig:Sample}a. 

-

Alternately, as-grown crystallites were ground into powder and then pressed under 5\,ton\,cm$^{-2}$ during 1\,min in order to obtain ceramic pellets of 13\,mm diameter and typically 1.8\,mm thick, in order to fabricate polycrystalline samples. All this was performed under controlled water-vapor-free atmosphere. Then gold was evaporated on both sides of the pellets, and finally a gold foil was pressed against both surfaces in order to ensure optimal electrical and acoustical contact. Silicone oil was used to optimize the latter. The structure of the polycrystalline samples is depicted in Figure~\ref{fig:Sample}b. Since RTO is known to be sensitive to humidity, the pellets were kept inside anhydrous silicone oil between the measurements.

\subsection{Charge distribution measurements}

\begin{figure}
\centering{}
\figfont
\psfrag{a}[Br][Br]{(a)}
\psfrag{Sample}[Bc][Bc]{Sample}
\psfrag{Vapp}[Bc][Bc]{$V_{app}$}
\psfrag{40dB}[Bl][Bl]{40\,dB}
\psfrag{Vm}[Bl][Bl]{$v_m$}
\psfrag{b}[Br][Br]{(b)}
\psfrag{Rho}[Br][Br]{$\rho$}
\psfrag{E}[Br][Br]{$E$}
\psfrag{S}[Br][Br]{$v_m$}
\psfrag{t=x/v}[Br][Br]{$t=x/v_s$}
\includegraphics[width=0.8\linewidth]{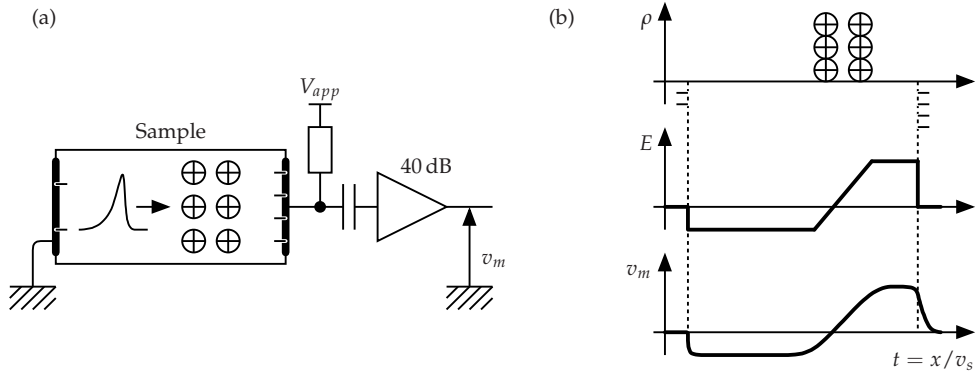}
\caption{Principle of the pressure-wave-propagation method  \cite{HoleJASA98}. (a) A continuous
voltage $V_{app}$ is applied across the sample. A pulsed pressure
wave travelling from the left to the right provokes a displacement
of the charges inside the sample, which generates in open-circuit
conditions a voltage signal as function of time, the shape of which is an
image of the electric field distribution inside the sample. (b) From top to bottom, charge distribution, electric field distribution, corresponding signal in open circuit conditions.} \label{fig:Setup}
\end{figure}

In the
pressure-wave-propagation method, a short-duration pressure pulse
is transmitted to the studied structure \cite{HoleJASA98}. The pressure wave is created by a high power piezogenerator [ref]. A 20 mm waveguide is used to decouple the signal from the pressure pulse generator excitation. The principle of the measurement
method is illustrated in Figure~\ref{fig:Setup}. As the
pressure wave propagates inside the sample, the charges encountered are slightly displaced
which induces a current in the measurement circuit connecting the
electrodes in short-circuit conditions, or a voltage in open-circuit
conditions. The obtained signal represents an image of the charge
distribution in short-circuit conditions, and of the internal electric
field in open-circuit conditions, time and position inside the sample
being connected by the velocity of sound v$_s$.

\begin{figure}
\centering{}
\figfont
\psfrag{a}[Br][Br]{(a)}
\psfrag{2000ns}[Bc][Bc]{2\,$\mu$s}
\psfrag{50ns/div1}[Br][Br]{50\,ns/div}
\psfrag{Pressure}[Bc][Bc]{Pressure}
\psfrag{Anode}[Br][Br]{Anode}
\psfrag{Cathode}[Br][Br]{Cathode}
\psfrag{Entrance1}[Bl][Bl]{Entrance}
\psfrag{Exit1}[Br][Br]{Exit}
\psfrag{Negative charges}[Bl][Bl]{Negative charges}
\psfrag{Positive charges}[Bl][Bl]{Positive charges}
\psfrag{Electric field [a.u.]}[Br][Br]{$\sim$Field (a.u.)}
\psfrag{b}[Br][Br]{(b)}
\psfrag{50ns/div2}[Bl][Bl]{50\,ns/div}
\psfrag{Entrance2}[Br][Br]{Entrance}
\psfrag{Exit2}[Bl][Bl]{Exit}
\psfrag{Ohmic contact}[Bc][Bc]{Zero-field region}
\includegraphics[width=0.5\linewidth]{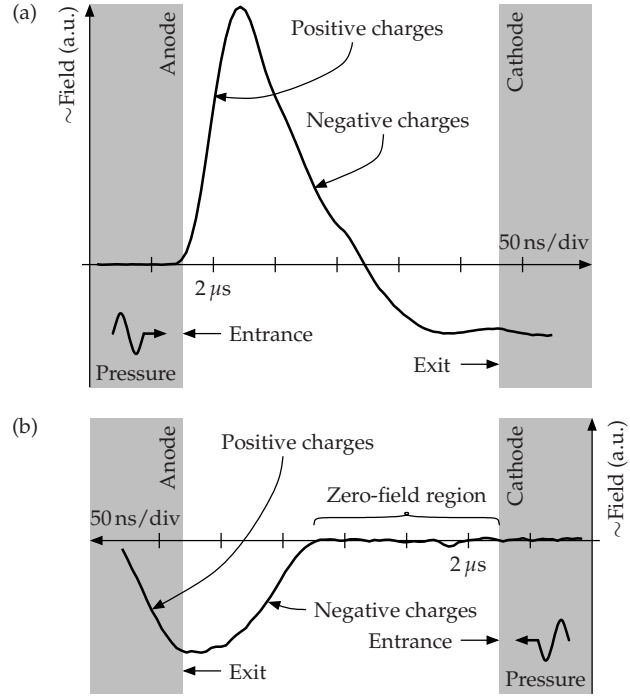}
\caption{Electric field distribution inside a single crystal sample (See Figure~\ref{fig:Sample}a). The electrodes are materialized by the grey boxes.(a) Measurement when the pressure pulse enters through the anode. Electric field is clearly non-zero at the anode. (b) Measurement when the pressure pulse enters through the cathode. No electric field is detected at the cathode and well inside the material.} \label{fig:Comparison}
\end{figure}

All measurements were carried out at room temperature and hygrometry and atmospheric
pressure. A typical signal is shown in Figure~\ref{fig:Comparison} for a
single crystal sample.
Piezoelectric transducers were coupled to the samples in preliminary
measurements to determine the exact position of the electrodes in
each sample. It was determined that at $t=1970$~ns in Figure~\ref{fig:Comparison}a,
the pressure pulse enters the sample. Because of the large sample
capacitance, measurements were performed in open-circuit conditions.
Therefore the first peak around $t=2\,\mu\mbox{s}$ is proportional
to the electric field at the interface. The rising edge is due to
positive charges on the anode electrode and the falling edge is due
to the presence of negative charges in the crystal. Such behavior is not visible on the other interface (cathode).

The presence of the epoxy and the complex geometry of the sample make it difficult to ensure a perfect pressure pulse propagation inside the single crystal sample. Thus a reliable analysis of the signal after
the first peak can not be garanteed. Therefore  the single-crystal sample was then mounted the opposite way in the sample
holder and the voltage was reversed. As a consequence, anode and cathode
remain the same with respect to the sample, but the pressure wave
enters from the cathode instead of the anode (see Figure~\ref{fig:Comparison}), in such a way that the integrity of the pressure pulse is now preserved at the cathode.
Surprisingly enough, it is clear from Figure~\ref{fig:Comparison}b
that when the pressure wave enters the sample from the cathode, almost
no interface peak is visible. On the contrary, the signal begins to rise later
indicating that up to about the middle of the sample, there is no
detectable electric field on the cathode side. The absence of electric
field on the cathode side indicates that there is no charge buildup which
implies in turn a substantial charge transfer across that interface,
contrarily to what happens at the anode interface where charges are
accumulated. Moreover, since the zero electric field extends up to the middle
of the sample, overall charge neutrality can be inferred in that region. On
the anode side, a large amount of negative charges (negative slope in
the signal) are clearly accumulated inside the RTO
crystal in the vicinity of the anode, with positive charge (positive slope) on the surface of the anode.

\begin{figure}
\centering{}
\figfont
\psfrag{a}[Br][Br]{(a)}
\psfrag{3100ns}[Bc][Bc]{3.1\,$\mu$s}
\psfrag{50ns/div}[Br][Br]{50\,ns/div}
\psfrag{Anode}[Br][Br]{Anode}
\psfrag{Cathode}[Br][Br]{Cathode}
\psfrag{Electric field [a.u.]}[Br][Br]{$\sim$Field (a.u.)}
\psfrag{Ohmic contact}[Bc][Bc]{Zero-field region}
\psfrag{b}[Br][Br]{(b)}
\psfrag{1min}[Bl][Bl]{1\,min}
\psfrag{5h}[Bl][Bl]{5\,h}
\includegraphics[width=0.5\linewidth]{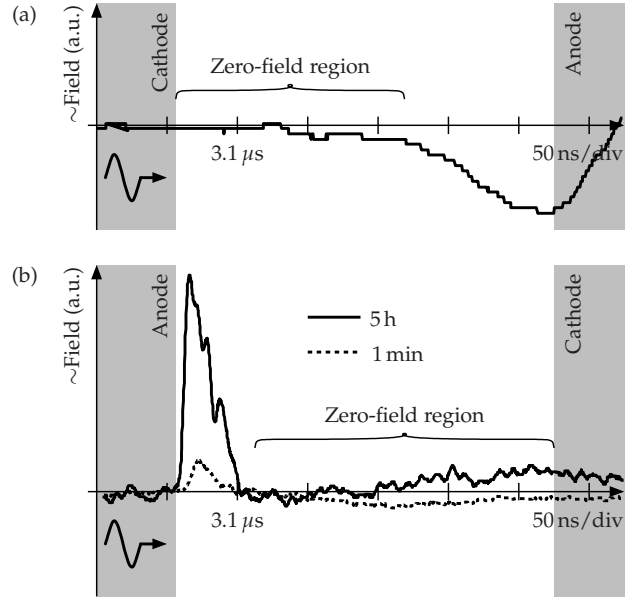}
\caption{Electric field distribution obtained in a polycrystalline sample (See Figure~\ref{fig:Sample}b). The electrodes are materialized by the grey boxes. (a) After applying a constant voltage of $+20$\,V during 5\,h. (b) 1\,min (dotted line) and 5\,h (solid line) after commuting the voltage from $+20$\,V to $-20$\,V. The electric field is rapidly decreased at the former anode and rapidly builds up in the vicinity of the new anode.} \label{ceram}
\end{figure}

In order to further confirm this surprising behavior by avoiding any parasitic signal due to the presence of epoxy resin and to understand the previous observation of colossal capacitance also in polycrystalline samples, the same measurements were performed in the polycrystalline samples prepared as described above. The results are presented in Figure \ref{ceram} for three consecutive measurements on the same sample.
The first observation, which comes as a surprise, is that the distribution of electric field is similar in a single crystal and a polycrystalline sample.  In Figure \ref{ceram}a is plotted the output signal after applying 20V to the sample during 5\,h.  Due to the absence of epoxy here, the positions of the input and output interfaces are easily determined by additional measurements during which first the sample and then the outcome contact are replaced by piezoelectrics. The electrodes are materialized by the gray boxes.  Again here, negative charges are accumulated in the vicinity of the anode inside the material while the electric field is null at the cathode and well inside the sample on the cathode side.  Figure \ref{ceram}b displays the result of the same measurement taken 1\,min after the voltage is reversed to $-20$\,V, which shows a rapid decrease of the electric field, and after 5h, where the signal is fully reversed with respect to Figure~\ref{ceram}a. It is noteworthy that the expected dispersion of sound in the polycrystallline material is consistent with the decreasing signal resolution with depth. 

\section{Results and discussion}

The first outcome of the measurements is to unveil the sign of
the ionic charge carriers without ambiguity. The ionic mobile species
are negatively charged, and there are no detectable accumulated mobile cations. This comes at odd with previous assumptions \cite{FedericciPRM2017,FedericciACSB2017} where mobile carriers were hypothesized to be oxygen vacancies,
and implies that the colossal capacitance takes place only at the
anode. For this to occur, the charge transfer at the anode has to
be almost zero. Possible candidates for mobile ions are $\mbox{O}^{2-}$
ions or $\mbox{OH}^{-}$ ions coming from potential water contamination. The material is indeed found to be strongly hygroscopic \cite{FedericciACSB2017} and bears similarities with vanadate structures \cite{chernova_layered_2009} and in particular vanadium pentoxide gels \cite{Livage:1991a,livage_sol-gel_1991,livage_sol-gel_1992,wang_synthesis_2006} or with hydrated ruthenium oxides \cite{zheng_new_1995}.
This remains to be further investigated, as well as the full evolution of the distribution of electric field as function of time under various thermal, hygroscopic and electric conditions.

This accumulation of charge is a further confirmation of the predominant electrostatic nature of the giant capacitance. As a matter of fact, in the case of electrochemical capacitance, as described and modelled by Jamnik and Maier \cite{Jamnik_2001}, the system would be locally electroneutral, and the pressure pulse propagation method would not reveal any charge accumulation. On the contrary, our measurements clearly demonstrate negative charges inside the RTO on the anode side and positive charges on the anode itself. The integral of the corresponding electric field obtained by calibrating the signal using a reference sample is indeed shown to correspond within 5\% to the applied voltage, leaving very little margin for electrochemical capacitance.

The second consequence is that the cathode interface is purely Ohmic, i.e.
there is a strong charge transfer at the interface and sufficient
conduction (either electronic or ionic) in the material near the cathode
to ensure a perfectly null electric field both on the cathode and
inside the crystal over sizable length, about typically 0.5~mm. Therefore the system behaves as if the cathode had
virtually shifted from the electrode to the middle of the sample.
The mechanisms for conduction close to the cathode could be either
that the departing ions have created electron or hole conduction,
or that the ionic conductivity is sufficient to ensure metallic-like
behavior in this region. Similar behavior has already been predicted
in crystals including titanium \cite{BaiatuJACS1990}. A possible mechanism is electron hopping between  Ti$^{3+}$ and Ti$^{4+}$ mixed valence cations leading to electron delocalization \cite{Livage:1991b}.
For reasons that are still to be understood, the cathode and the anode
interfaces thus behave quite differently in this system, with substantial
charge transfer at the cathode and immaterial charge transfer at the
anode. This is true both for Al electrodes used for the single-crystals and Au electrodes used for the ceramics, thus pointing to an intrinsic physical property of the material. This could be related to a strong electron/hole asymmetry for transfer processes. Although these processes are still to be elucidated as well as the
dominant mechanism for conduction near the cathode, the present observation
can be an explanation for the colossal equivalent permittivity observed
in RTO. Indeed, in contradiction to
conventional electrolytes where both charge polarities coexist, our findings indicate a charge compensation in the bulk on the cathode side, which enables much higher charge accumulation at the anode. 
In any case the mechanism at play has to be highly reversible since we did not observe degradation in the I/V curves after thousands of cycles.

Finally, the third remarkable observation is that single crystals and ceramics behave qualitatively the same way.  This rules out the possibility of accumulation of ions inside the grains, since in this case, an accumulation of charges should be seen also on the cathode side. This therefore tends to indicate that ionic motion is not only intragrain but also intergrain. In addition it opens the way to easier and cheaper possibilities for application in the domain of supercapacitors for energy storage.

\section{Conclusion}

We have shown from pressure-wave-propagation measurements
that negative ionic carriers accumulate in $\mbox{Rb}_{2}\mbox{Ti}_{2}\mbox{O}_{5}$ single crystals and polycrystalline samples.
When anions (possibly O$^{2-}$ or OH$^-$) accumulate at the anode, the cathode and the region
nearby experience an Ohmic-like behavior resulting in a zero electric field
in that region, thus leading to a "virtual cathode" extending deep inside the material. These findings potentially explain the colossal capacitance observed in $\mbox{Rb}_{2}\mbox{Ti}_{2}\mbox{O}_{5}$ as it enables much higher accumulated charge density. Further work is required to understand the charge transfer processes at the cathode, the dynamics of the charge relaxation as well as the nature of the mobile ions.

\section*{Acknowledgments}
N. Dragoe and D. Bérardan from ICMMO are gratefully acknowledged for fruitful discussions as well as for access to sample press under controlled atmosphere. One of us (RF) thanks Nexans France for support through the Chaire at ESPCI Paris. This work has received support under the program "Investissement d'Avenir"  launched by the French Government and implemented by ANR with the reference ANR-10-IDEX-0001-02 PSL.


\section*{References}
\begin{thebibliography}{10}
\expandafter\ifx\csname url\endcsname\relax
  \def\url#1{\texttt{#1}}\fi
\expandafter\ifx\csname urlprefix\endcsname\relax\def\urlprefix{URL }\fi
\expandafter\ifx\csname href\endcsname\relax
  \def\href#1#2{#2} \def\path#1{#1}\fi

\bibitem{SmithaJMS2005}
B.~Smitha, S.~Sridhar, A.~Khan,
  \href{https://doi.org/10.1016/j.memsci.2005.01.035}{Solid polymer electrolyte
  membranes for fuel cell applications - a review}, Journal of Membrane Science
  259 (2005) 10--26.
\newblock \href {http://dx.doi.org/10.1016/j.memsci.2005.01.035}
  {\path{doi:10.1016/j.memsci.2005.01.035}}.
\newline\urlprefix\url{https://doi.org/10.1016/j.memsci.2005.01.035}

\bibitem{VangariJEE2013}
M.~Vangari, T.~Pryor, L.~Jiang,
  \href{https://doi.org/10.1061/(ASCE)EY.1943-7897.0000102}{Supercapacitors:
  review of materials and fabrication methods}, Journal of Energy Engineering
  139 (2013) 72--79.
\newblock \href {http://dx.doi.org/10.1061/(ASCE)EY.1943-7897.0000102}
  {\path{doi:10.1061/(ASCE)EY.1943-7897.0000102}}.
\newline\urlprefix\url{https://doi.org/10.1061/(ASCE)EY.1943-7897.0000102}

\bibitem{LiAEM2015}
J.~Li, C.~Ma, M.~Chi, C.~Liang, N.~J. Dudney,
  \href{https://doi.org/10.1002/aenm.201401408}{Solid {E}lectrolyte: the {K}ey
  for {H}igh-{V}oltage {L}ithium {B}atteries}, Adv. Energy Mater. 5 (2015)
  1401408--1--6.
\newblock \href {http://dx.doi.org/10.1002/aenm.201401408}
  {\path{doi:10.1002/aenm.201401408}}.
\newline\urlprefix\url{https://doi.org/10.1002/aenm.201401408}

\bibitem{augustyn_pseudocapacitive_2014}
V.~Augustyn, P.~Simon, B.~Dunn,
  \href{http://xlink.rsc.org/?DOI=c3ee44164d}{Pseudocapacitive oxide materials
  for high-rate electrochemical energy storage}, Energy \& Environmental
  Science 7~(5) (2014) 1597.
\newblock \href {http://dx.doi.org/10.1039/c3ee44164d}
  {\path{doi:10.1039/c3ee44164d}}.
\newline\urlprefix\url{http://xlink.rsc.org/?DOI=c3ee44164d}

\bibitem{CarterTVT2012}
R.~Carter, A.~Cruden, P.~J. Hall,
  \href{https://doi.org/10.1109/TVT.2012.2188551}{Optimizing for {E}fficiency
  or {B}attery {L}ife in a {B}attery/{S}upercapacitor {E}lectric {V}ehicle},
  IEEE Transactions on Vehicular Technology 61 (2012) 1526--1533.
\newblock \href {http://dx.doi.org/10.1109/TVT.2012.2188551}
  {\path{doi:10.1109/TVT.2012.2188551}}.
\newline\urlprefix\url{https://doi.org/10.1109/TVT.2012.2188551}

\bibitem{ANDERSSON:1960fi}
S.~Andersson, A.~Wadsley, Five co-ordinated titanium in
  $\mbox{K}_2\mbox{Ti}_2\mbox{O}_5$, Nature (1960) 499.

\bibitem{Andersson:1961vf}
S.~Andersson, A.~Wadsley, The crystal structure of
  $\mbox{K}_2\mbox{Ti}_2\mbox{O}_5$, Acta Chem. Scand. 15 (1961) 663--669.

\bibitem{FedericciPRM2017}
R.~Federicci, S.~Hol\'e, A.~F. Popa, L.~Brohan, B.~Baptiste, S.~Mercone,
  B.~Leridon,
  \href{https://journals.aps.org/prmaterials/pdf/10.1103/PhysRevMaterials.1.032001}{{R}b$_2${T}i$_2${O}$_5$:
  {S}uperionic conductor with colossal dielectric constant}, Phys. Rev.
  Materials 1 (2017) 032001.
\newblock \href {http://dx.doi.org/10.1103/PhysRevMaterials.1.032001}
  {\path{doi:10.1103/PhysRevMaterials.1.032001}}.
\newline\urlprefix\url{https://journals.aps.org/prmaterials/pdf/10.1103/PhysRevMaterials.1.032001}

\bibitem{FedericciJAP2018}
R.~Federicci, S.~Holé, V.~Démery, B.~Leridon,
  \href{https://doi.org/10.1063/1.5036841}{Memory effects in the ion conductor
  {R}b$_2${T}i$_2${O}$_5$}, Journal of Applied Physics 124~(15) (2018) 152104.
\newblock \href {http://arxiv.org/abs/https://doi.org/10.1063/1.5036841}
  {\path{arXiv:https://doi.org/10.1063/1.5036841}}, \href
  {http://dx.doi.org/10.1063/1.5036841} {\path{doi:10.1063/1.5036841}}.
\newline\urlprefix\url{https://doi.org/10.1063/1.5036841}

\bibitem{LaurenceauPRL77}
P.~Laurenceau, G.~Dreyfus, J.~Lewiner,
  \href{https://doi.org/10.1103/PhysRevLett.38.46}{New principle for the
  determination of potential distributions in dielectrics}, Phys. Rev. Lett. 38
  (1977) 46--49.
\newblock \href {http://dx.doi.org/10.1103/PhysRevLett.38.46}
  {\path{doi:10.1103/PhysRevLett.38.46}}.
\newline\urlprefix\url{https://doi.org/10.1103/PhysRevLett.38.46}

\bibitem{HoleTDEI2003}
S.~Hol\'e, T.~Ditchi, J.~Lewiner,
  \href{https://doi.org/10.1109/TDEI.2003.1219652}{Non-destructive methods for
  space charge distribution measurements: what are the differences?}, IEEE
  Trans. Dielectr. EI. 10 (2003) 670--677.
\newblock \href {http://dx.doi.org/10.1109/TDEI.2003.1219652}
  {\path{doi:10.1109/TDEI.2003.1219652}}.
\newline\urlprefix\url{https://doi.org/10.1109/TDEI.2003.1219652}

\bibitem{FedericciACSB2017}
R.~Federicci, B.~Baptiste, F.~Finocchi, F.~Popa, L.~Brohan, K.~B\'eneut,
  P.~Giura, G.~Rousse, A.~Descamps-Mandine, T.~Douillard, A.~Shukla,
  B.~Leridon, \href{https://doi.org/10.1107/S2052520617013646}{The crystal
  structure of {R}b$_2${T}i$_2${O}$_5$}, Acta Crystallographica Section B 73
  (2017) 1142--1150.
\newblock \href {http://dx.doi.org/10.1107/S2052520617013646}
  {\path{doi:10.1107/S2052520617013646}}.
\newline\urlprefix\url{https://doi.org/10.1107/S2052520617013646}

\bibitem{HoleJASA98}
S.~Hol\'e, J.~Lewiner, \href{https://doi.org/10.1121/1.423863}{Design and
  optimization of unipolar pressure pulse generators with a single transducer},
  J. Acoust. Soc. Am. 104 (1998) 2790--2797.
\newblock \href {http://dx.doi.org/10.1121/1.423863}
  {\path{doi:10.1121/1.423863}}.
\newline\urlprefix\url{https://doi.org/10.1121/1.423863}

\bibitem{chernova_layered_2009}
N.~A. Chernova, M.~Roppolo, A.~C. Dillon, M.~S. Whittingham,
  \href{http://xlink.rsc.org/?DOI=b819629j}{Layered vanadium and molybdenum
  oxides: batteries and electrochromics}, Journal of Materials Chemistry
  19~(17) (2009) 2526.
\newblock \href {http://dx.doi.org/10.1039/b819629j}
  {\path{doi:10.1039/b819629j}}.
\newline\urlprefix\url{http://xlink.rsc.org/?DOI=b819629j}

\bibitem{Livage:1991a}
J.~Livage, Vanadium pentoxide gels, Chem. Mater., 3 (1991) 578.

\bibitem{livage_sol-gel_1991}
J.~Livage, \href{http://doi.org/10.1016/0167-2738(92)90234-G}{Sol-gel ionics},
  Solid State Ionics (1992) 307--313\href
  {http://dx.doi.org/10.1016/0167-2738(92)90234-G}
  {\path{doi:10.1016/0167-2738(92)90234-G}}.
\newline\urlprefix\url{http://doi.org/10.1016/0167-2738(92)90234-G}

\bibitem{livage_sol-gel_1992}
J.~Livage, \href{https://doi.org/10.1016/0167-2738(96)00336-0}{Sol-gel
  chemistry and electrochemical properties of vanadium oxide gels}, Solid State
  Ionics (1996) 935--942\href {http://dx.doi.org/10.1016/0167-2738(96)00336-0}
  {\path{doi:10.1016/0167-2738(96)00336-0}}.
\newline\urlprefix\url{https://doi.org/10.1016/0167-2738(96)00336-0}

\bibitem{wang_synthesis_2006}
Y.~Wang, G.~Cao, \href{http://pubs.acs.org/doi/abs/10.1021/cm052765h}{Synthesis
  and {Enhanced} {Intercalation} {Properties} of {Nanostructured} {Vanadium}
  {Oxides}}, Chemistry of Materials 18~(12) (2006) 2787--2804.
\newblock \href {http://dx.doi.org/10.1021/cm052765h}
  {\path{doi:10.1021/cm052765h}}.
\newline\urlprefix\url{http://pubs.acs.org/doi/abs/10.1021/cm052765h}

\bibitem{zheng_new_1995}
J.~P. Zheng, \href{http://jes.ecsdl.org/cgi/doi/10.1149/1.2043984}{A {New}
  {Charge} {Storage} {Mechanism} for {Electrochemical} {Capacitors}}, Journal
  of The Electrochemical Society 142~(1) (1995) L6.
\newblock \href {http://dx.doi.org/10.1149/1.2043984}
  {\path{doi:10.1149/1.2043984}}.
\newline\urlprefix\url{http://jes.ecsdl.org/cgi/doi/10.1149/1.2043984}

\bibitem{Jamnik_2001}
J.~Jamnik, J.~Maier, \href{http://xlink.rsc.org/?DOI=b100180i}{Generalised
  equivalent circuits for mass and charge transport: chemical capacitance and
  its implications}, Physical Chemistry Chemical Physics 3~(9) (2001)
  1668--1678.
\newblock \href {http://dx.doi.org/10.1039/b100180i}
  {\path{doi:10.1039/b100180i}}.
\newline\urlprefix\url{http://xlink.rsc.org/?DOI=b100180i}

\bibitem{BaiatuJACS1990}
T.~Baiatu, R.~Waser, K.-H. H\"ardtl,
  \href{https://doi.org/10.1111/j.1151-2916.1990.tb09811.x}{{d}c {E}lectrical
  {D}egradation of {P}erovskite-{T}ype {T}itanates: {III}, {A} {M}odel of the
  {M}echanism}, Journal of the American Ceramic Society 73 (1990) 1663--1673.
\newblock \href {http://dx.doi.org/10.1111/j.1151-2916.1990.tb09811.x}
  {\path{doi:10.1111/j.1151-2916.1990.tb09811.x}}.
\newline\urlprefix\url{https://doi.org/10.1111/j.1151-2916.1990.tb09811.x}

\bibitem{Livage:1991b}
J.~Livage, Interface properties of vanadium pentoxide gels, Mat. Res. Bull 26
  (1991) 1173.

\end{thebibliography}

\end{document}